\documentstyle[12pt]{article}
\def\beq{\begin{equation}}
\def\eeq{\end{equation}}
\def\bea{\begin{eqnarray}}
\def\eea{\end{eqnarray}}
\def\bq{\begin{quote}}
\def\eq{\end{quote}}

\def\nnb{\nonumber}
\def\ga{\left(}
\def\dr{\right)}
\def\aga{\left\{}
\def\adr{\right\}}

\def\rar{\rightarrow}
\def\nnb{\nonumber}

\begin{document}
\topmargin -2.0cm
\oddsidemargin +0.2cm
\evensidemargin -1.0cm
\pagestyle{empty}
\begin{flushright}
{CERN-TH.7103/93}\\
PM 93/45
\end{flushright}
\vspace*{5mm}
\begin{center}
{{\bf
 $\Upsilon \bar BB$ COUPLINGS, SLOPE OF THE ISGUR-WISE
FUNCTION \\ AND IMPROVED ESTIMATE OF \boldmath{$V_{cb}$}}} \\
\vspace*{1.5cm}
{\bf S. Narison} \\
\vspace{0.3cm}
Theoretical Physics Division, CERN\\
CH - 1211 Geneva 23\\
and\\
Laboratoire de Physique Math\'ematique\\
Universit\'e de Montpellier II\\
Place Eug\`ene Bataillon\\
34095 - Montpellier Cedex 05\\
\vspace*{2.0cm}
{\bf ABSTRACT} \\ \end{center}
\vspace*{2mm}
\noindent
We estimate the sum of the $\Upsilon \bar BB $ couplings using QCD
Spectral Sum Rules (QSSR).
Our result implies  the phenomenological bound  $\xi'(vv'=1) \geq
-1.04$
for the slope of the Isgur-Wise function. An analytic estimate
of the (physical) slope to two loops
 within QSSR leads to the accurate
 value $\xi'(vv'=1) \simeq -(1.00
\pm 0.02)$ due to the (almost) complete
 cancellations between the perturbative and non-perturbative corrections
at the stability points.
Then, we deduce, from the present data, the improved estimate
$\vert V_{cb} \vert  \simeq \ga 1.48 \mbox{ps}/\tau_B
\dr ^{1/2}(37.3 \pm 1.2 \pm 1.4)\times 10^{-3}$ where the first error
comes from the data analysis and the second one from the different
model parametrizations of the Isgur-Wise function.

\vspace*{2.5cm}
\noindent

\vspace*{1.5cm}

\begin{flushleft}
CERN-TH.7103/93 \\
PM 93/45\\
January 1994
\end{flushleft}
\vfill\eject
\pagestyle{empty}

\setcounter{page}{1}
\pagestyle{plain}

\section{Introduction}
With the advent of the heavy quark symmetry \cite{MAN},
there has been
considerable interest and progress in the understanding of the
semileptonic
form factors of the transition of a heavy quark
into another heavy quark,
as in this infinite mass limit
all semileptonic form factors reduce to the single Isgur-Wise
function \cite{IWI}.
The question is whether this result in the heavy quark
infinite mass limit can be applied to the physical B and D mesons.
Progress has been made for including the $1/M_Q$ corrections. At the
non-recoil point, the  $1/M_Q$ terms cancel in the
form factors
\cite{LUK}, while for the
leptonic decay constant, one expects a large $1/M_Q$ correction
as indicated by lattice \cite{SAC}
and QCD Spectral Sum Rules (QSSR)\cite{SN1} estimates.
QCD vertex sum rules have also been applied for evaluating the
semi-leptonic form factors at $q^2=0$ \cite{OVI}--\cite{DOS}
and the Isgur-Wise function
at $q^2_{max}$\cite{DOS}--\cite{NEU}.
Although the QSSR results are quite impressive
and the authors \cite{DOS}--\cite{NEU}
expect that working with the ratio of the vertex
over the two-point pseudoscalar sum rules eliminates different systematic
uncertainties, one should not forget that, in this derivation,
certain assumptions on the choice
of the QSSR scales and QCD continuum thresholds, which introduce
uncertainties, have to be done.
Most of these choices are based on
the model discussed in Ref. \cite{BLO}.
More rigorous and less
model dependent is the bound obtained recently by \cite{TAR2},
where one  exploits
the analyticity and the unitarity of the {\it b}-number form factor
in the limit $M_c=M_b$ (see also \cite{DOM},\cite{TAR1}).
The bound on the slope of the form factor in \cite{TAR2}
depends on the residues and on the positions of $\Upsilon$ poles.
The most rigorous bound comes from the normalization
condition $F(0)=1$ of the {\it b}-number form factor leading to
$F'(v.v' =1)
\ge -6$,
while the inclusion of
$\Upsilon \bar BB$
couplings, assumed to be of order one each,
leads to phenomenological bound \cite{TAR2}:
\beq
F'(v.v' =1)
\ge -1.5,
\eeq
which is 4 times much stronger than the previous one.

The purpose of this note is to present an estimate of the
$\Upsilon \bar BB$ couplings which play an important role in the
phenomenological derivation of the previous bounds
\cite{TAR2}, \cite{DOM}.
We shall also estimate directly this slope from
analytic expressions to two-loops in
the QSSR approach. Then, we shall deduce an estimate of the CKM mixing
angle $V_{cb}$.

As
the $\bar BB$ pairs are below the $\Upsilon$(1S,2S,3S)
states, there are
no available data
for estimating such couplings.
The only experimental available information concerns the
coupling of the $\Upsilon$(4S) which lies above the $\bar BB$
threshold.  From the data of
its leptonic and total widths, one can deduce
\cite{TAR2} :
\beq
|\eta_{\Upsilon_4}|
 \equiv \frac{g_{{\Upsilon}_4 \bar BB}}{2\gamma_{\Upsilon_4}}
 \leq 0.75 \pm 0.15,
\eeq
where
\beq
\Gamma ({\Upsilon}(4S) \rightarrow e^+e^-)=  \frac{1}{3}\pi\alpha^2
\ga \frac{1}{3} \dr ^2 \frac{M_{\Upsilon_4}}{ \gamma^2_{\Upsilon_4}}
\eeq
and
\beq
\Gamma \ga\Upsilon(4S) \rightarrow \bar BB \dr
=M_{{\Upsilon}_4}
\ga\frac{ g^2_{{\Upsilon_4}\bar BB}}{48\pi} \dr
\ga 1-\frac{4M^2_B}{M^2_{\Upsilon_4}} \dr ^{3/2},
\eeq
with the normalization:
\beq
<0|-\bar b \gamma^{\mu}b|\Upsilon_i> = \epsilon^{\mu}
\frac{ M^2_{\Upsilon_i}}{2\gamma_{\Upsilon_i}},
\eeq
where we have included in the definition of the current the negative
sign related to the {\it b}-quark charge.
In principle, the  couplings of the
three lightest states are unconstrained. However,  using the location
of the three $\Upsilon$ poles and the fact that the b-number form factor
of the B-meson is 1 at $q^2=0$, Ref.\cite{TAR2} obtains
the following bounds:
\beq
|\eta_{\Upsilon_1}| \leq 9.9 \times 10^3 \ \ , \ \
|\eta_{\Upsilon_2}| \leq 17.1 \times 10^3 \ \ , \ \
|\eta_{\Upsilon_3}| \leq 8.1 \times 10^3,
\eeq
which, as already pointed by the authors, are
quite weak compared to the one for $\eta_4$
obtained previously from the data.
\section{QSSR estimate of the $\Upsilon \bar BB$ coupling}
In so doing, we consider the vertex correlator :
\bea
V^{\mu}(q^2,p^2,p'^2) =  (i)^2 \int d^4x \ d^4y e^{i(px-p'y)}
<0| \mbox{T}J_5(x)J^{\mu}(0)J^{\dagger}_5(y)|0> \nonumber \\
\equiv \ga p+p' \dr ^\mu V_+(q^2,p^2,p'^2),
\eea
built with the local quark currents:
\beq
J_5(x) = (m_d+M_b) : \bar d(i\gamma_5)b : \ \ , \ \
J^\mu  = \ -: \bar b\gamma^\mu b :.
\eeq
$V_+$ obeys the double dispersion relation :
\beq
V_+(q^2,p^2,p'^2) = -\frac{1}{\pi^2}
\int_{M^2_b}^{\infty} \frac {dt}{t-p^2}
 \int_{M^2_b}^{\infty} \frac {dt'}{t'-p'^2} \mbox{Im}
  V_+(q^2,t,t') +...,
\eeq
where ... means polynomial subtraction constants and $q^2 \equiv
(p-p')^2 \leq 0$.
For the estimate of the $\sum \eta_{\Upsilon_i}$
couplings, we evaluate the correlator
at $q^2$ = 0 similarly to the estimate of the form factors of the B
semi-leptonic decays using vertex sum rules. We shall limit ourselves
to the lowest order contribution in $\alpha_s$ but include the
non-perturbative condensate contributions in the OPE.
This approximation
has also
given a quite good prediction in different estimates of the
B and D decay form factors.
So we, a priori, expect that a similar feature will
hold in our analysis. The QCD expression of the three-point function
has been evaluated in the literature \cite{OVI}, including higher
dimension condensates. Our case corresponds to putting $M_b=M_c$ in
this paper. Therefore, the perturbative contribution reads, to leading
order in $\alpha_s$:
\beq
\mbox{Im} V_+^{pert}(q^2,t,t') \simeq -\frac{3}{4}q^2
\frac{\ga M_b^4-tt' \dr}{\ga \ga t+t' -q^2 \dr ^2 -4 tt'
\dr ^{3/2}},
\eeq
which shows that at $q^2=0$, the lowest
order perturbative contribution to the spectral function
vanishes. We have also checked this result from a direct evaluation
of the triangle perturbative diagram. Therefore, in this case,
the leading contribution
comes from the light quark condensate and reads :
\beq
V_+(0,p^2,p'^2) = M^3_b<\bar dd> \frac{1}{(p^2-M^2_b)(p'^2-M^2_b)},
\eeq
which is not the case  of the other B decay form factors
studied previously at $q^2=0$. So, from this particular feature,
we (intuitively) expect that $V_+(0)$
is much smaller than the previous form factors. We parametrize
the phenomenological side of the vertex by introducing the B and
$\Upsilon$ couplings via:
\bea
<0|J_5|B> = \sqrt 2 f_BM^2_B,  \nonumber \\
<0|J^{\mu}|\Upsilon_i> = \epsilon^{\mu}
\frac{ M^2_{\Upsilon_i}}{2\gamma_{\Upsilon_i}}
\eea
and we insert the intermediate states in (7).
Using the definition of the $\Upsilon \bar BB$ coupling in (2), we
have at $q^2=0$:
\beq
V_+(0,p^2,p'^2) \simeq  -\frac{2f^2_BM^4_B}{(p^2-M^2_B)(p'^2-M^2_B)}
\sum_i {\eta_{\Upsilon_i}}.
\eeq
If one uses the quark hadron (semi)local
duality picture by simply equating the
phenomenological and QCD sides of the vertex, one obtains the sum
rule:
\beq
\sum_i {\eta_{\Upsilon_i}} \simeq \ga \frac{M_b}{M_B} \dr ^3
\frac{<\bar dd>}{2f^2_BM_B}.
\eeq
Using for $f_B$ the local duality constraint \cite{ZAL}:
\beq
2f^2_B M_B\simeq \frac{1}{\pi^2}\ga E^B_c \dr ^3
\ga \frac{M_b}{M_B} \dr ^3,
\eeq
where $E^B_c \simeq 1.3$ GeV \cite{SN1}
is the B continuum energy, one obtains :
\beq
\sum_i {\eta_{\Upsilon_i}}\simeq  \frac{\pi^2}{\ga E^B_c \dr^3}
 < \bar dd > \simeq -0.07,
\eeq
if one uses $ < \bar dd > (E^B_c)
 \simeq$ -(250MeV)$^3$. The previous
constraint indicates that $\eta_{\Upsilon_i}$
remains constant for $M_b \rightarrow
\infty$, in agreement with the
expectation from the ``na\"{\i}ve $M_b$ counting rule'':
\beq
g_{\Upsilon \bar BB}\rar M^{1/2}_b  \ \ , \ \
\gamma_\Upsilon\rar M^{1/2}_b.
\eeq
This feature increases our confidence on the physical meaning of the
constraint in (14). Moreover, the previous constraint also indicates
that the sum of couplings in (16) is almost independent of the b-quark
mass as we shall check later on in the case of the J/$\psi$.

One can improve further the previous constraint by working with the
Laplace (Borel) sum rules and by including the contribution of the
quark-gluon mixed condensate $g<\bar d \sigma^{\mu\nu}
\frac{\lambda^a}{2} G^a_{\mu\nu} d> \equiv M^2_0 <\bar dd>
$. In this way, the relativistic Laplace operator leads to the change :
\beq
\frac{1}{p^2-M_P^2}\rar \tau e^{-\tau M_P^2} \ \ , \ \
\frac{1}{p'^2-M_P^2}\rar \tau' e^{-\tau' M_P^2},
\eeq
where $\tau$ (resp. $\tau'$) are the sum rule variables associated
to $p^2$ (resp. $p'^2$). For convenience and
because of the
symmetry of the vertex,
we shall take the natural choice
$\tau$ =
$\tau'$ in our analysis.
However,
this choice does not have any noticeable
effect in our conclusion.
We use the following values
of the QCD parameters \cite{SN1},\cite{SN5}:
\bea
f_B \simeq (1.59 \pm 0.26)f_\pi \nonumber \\
M_b(p^2=M^2_b) \simeq (4.59 \pm
0.05) \, \mbox{GeV} \nonumber \\
<\bar dd> = -((189 \pm 7)
\ \mbox{MeV})^3 \ga -\log(\tau^{-1/2} \Lambda) \dr
^{-2/\beta_1}  \nonumber \\
M^2_0 \simeq (0.80 \pm 0.10) \,  \mbox{GeV}^2,
\eea
where $-\beta_1 =\frac{1}{2}(11-\frac{2n}{3})$ for $SU(n)_f$ and
$\Lambda = 260 \pm 50$ MeV.
We give the result for $\sum \eta_{\Upsilon_i} $ in Fig. 1
versus the sum rule
variable $\tau$. Due to the absence of the perturbative contribution
in our leading order analysis, the result is insensitive to the
continuum contribution. The stability in $\tau$ is reached at 0.2 GeV$^{-
2}$, a value that is quite similar to the one where $f_B$ is optimal
\cite{SN1}.
At the stability
point, one obtains:
\beq
\sum_i \eta_{\Upsilon_i} \simeq -0.224,
\eeq
where the error is negligible (0.004), if one uses the correlated
values of $f_B$ and $M_b$. We estimate the maximal error by taking the
uncorrelated values of the previous two parameters. Then, we
obtain, at the order we are working, the estimate :
\beq
\sum_i \eta_{\Upsilon_i} \simeq -(0.224  \pm 0.064).
\eeq
An improvement of this result needs an evaluation of the
radiative correction to the vertex function. However, we do not
expect that the higher order terms will modify the present
leading order
results by more than
a factor 2-3, if no marginal terms break the conventional
OPE or/and if there are no anomalous couplings that drastically
modify the
parametrization of the spectral function in (13).

The result from the simple local duality relation
in (16)
corresponds to the case where $\tau \rar 0$. The difference between
the local duality and Laplace sum rules results can indicate the
possible large role of the continuum for $\tau \rar 0$ which is
negligible in the Laplace sum rule analysis.
\par
\noindent
We test the quality of our estimate by
applying the method in the J/$\psi$ channel where the couplings
of the 3S and 4S states are also bounded experimentally to be :
\beq
\eta_{\psi_3} \leq 0.47 \ , \  \eta_{\psi_4} \leq 0.17.
\eeq
As we have discussed in the case of the decay constant \cite{ZAL}, the
simple duality constraint
in (15) reproduces correctly the ratio $f_B/f_D$, though its prediction
for the absolute value is inaccurate. Then, we also expect
that the ratio of the coupling $\eta_i$ is well reproduced by (16). It
gives :
\beq
\sum_i {\eta_{\psi_i}} \simeq -\ga \frac{E_c^B}{E_c^D} \dr^3
\sum_i {\eta_{\Upsilon_i}} \simeq 0.40,
\eeq
where we have used $E^D_c \simeq 1.08$ GeV \cite{SN1},\cite{ZAL}.
A direct Laplace
(Borel) sum rule estimate anologous to the one used to get (21) gives
an optimization scale $\tau \simeq $ 0.8 GeV$^{-2}$, very similar to the
one for $f_D$ \cite{SN1}. Using
$f_D \simeq (1.31 \pm 0.12)f_\pi $
and
$M_c(p^2=M^2_c) \simeq (1.45 \pm
0.05) \mbox{GeV}$
\cite{SN1},
the sum rule
gives the estimate :
\beq
\sum_i {\eta_{\psi_i}} \simeq  0.34 \pm 0.02.
\eeq
This result
compares quite well with the previous leading order estimate from
local duality sum rule in (23) and with the
experimental bounds in (22). It also indicates that the sum of couplings
is almost independent of the heavy quark mass, which is manifest in the
local duality constraints.
This test increases further our confidence on the
numbers obtained in (21).
\par
\noindent
Moreover, our result applies if
the quark structure of the
vertex remains valid for this particular process but
cannot
be used if
the  $\bar BB, \bar {B^*}B^*$ pairs  are
molecules
formed by Van Der Vaals
like forces. In this case a vertex sum rule approach with four-quark
currents
similar to the one
done for the $K\bar K$ molecule \cite{SN12} becomes more adequate.

\section{Bound on the slope of the Isgur-Wise function}
We have estimated the sum of the $\Upsilon \bar BB$ couplings using
a conventional vertex sum rules analysis within the quark
structure of the $\bar BB$ meson pairs.
Despite the leading order approximation that
we have used for deriving the values of the couplings, we expect
that our results are valid within a factor of 2-3,
which is a conservative
estimate of the radiative corrections not included here.
A comparison of the result
with the experimental bound in (2) indicates that the
bound for the $\Upsilon(4S)$ coupling is satisfied by the
sum of the couplings of the different $\Upsilon$ states. In order to
see the effects of our results for the bound on the slope of the
Isgur-Wise function as derived in \cite{TAR2}, we shall consider the
following different scenarios on the strength of each coupling given
the constraint in (19) for the sum:
The first scenario, which seems to be the most plausible
phenomenologically, due to the experimental suppression of the electronic
widths of the 2S and 3S states,
is the one where the coupling of the 1S state is
much larger than the previous ones (vector meson dominance) (first row
in the tables). The second and third scenarios (second and third rows)
are the ones where the absolute values of the couplings are equal, with
also the possibility to have a cancellation between the couplings of
the 2S and 3S. The fourth possibility (fourth row)
is the one where the coupling of
the 3S almost saturates the experimental bound for the 4S given in (1).
The
one of the 2S is assumed to be about the meson mass squared ratio
$(M_3/M_2)^2$ times the one of the 2S.

The upper and lower bounds on the slope of the
{\it b}-number form factor
are given in these tables \cite{TAR3}. One can notice
that increasing the sum of couplings by a factor 3 only affects
the bounds by $15\%$. From the tables, one can deduce the
conservative phenomenological bounds for the slope of the {\it b}-number
form factor:
\beq
-(0.88\sim 1.34) \leq F'(vv'=1) \leq (0.08\sim 0.52).
\eeq
The lower bound is comparable with and even slightly stronger than the
conservative bound in (1)  given by Ref. \cite{TAR2}, while the upper
bound is weaker than the Bjorken bound of -1/4
\cite{BJO}.
Using its relation with the slope of the IW function $\xi'$ \cite{GOR}:
\beq
\xi'(1) \simeq  F'(1) - \frac{16}{75} \log \alpha_s (M_b),
\eeq
we can deduce the conservative lower bound:
\beq
\xi'(vv'=1) \geq -1.04.
\eeq
The previous results
question the
accuracy of the experimental domain \cite{ARG}:
\beq
 -2.3 \leq \xi'(vv'=1) \leq -1.17,
\eeq
obtained after extrapolating the
data until the non-recoil point.
Our results also indicate
that the smallness of the sum of the $\Upsilon
\bar BB $ couplings, and presumably, of each individual coupling
derived in this paper, raises again some doubts on the accuracy
of existing models and methods
used for determining the mixing angle V$_{cb}$.

\begin{table}
    \begin{center}
%
%
%
\begin{tabular}{||c|c|c|c|c||}
   \hline
   $\eta_1$ & $ \eta_2  $ &
    $\eta_3  $ & $ F'(1)_{lower}   $ & $ F'(1)_{upper} $ \\
\hline
-0.224 & 0.000      & 0.000       & -0.967 & 0.447 \\
\hline
-0.224 & -0.224 &  0.224 & -0.882 & 0.516 \\
\hline
-0.080 & -0.080 & -0.080 & -0.950 & 0.465 \\
\hline
-1.760 & +0.890 &  0.720 & -1.183 & 0.235 \\
\hline
\end{tabular}
\end{center}
\caption{ \label{X.1}
 Upper and lower bounds for the slope of the {\it b}-number form
factor
for various phenomenological values of the couplings $\eta_i$
with $\sum \eta_i =-0.224$.}
\end{table}

\begin{table}
    \begin{center}
%
%
%
\begin{tabular}{||c|c|c|c|c||}
   \hline
   $\eta_1$ & $ \eta_2  $ &
    $\eta_3  $ & $ F'(1)_{lower}   $ & $ F'(1)_{upper} $ \\
\hline
-0.672 & 0.000      & 0.000       & -1.118 & 0.306 \\
\hline
-0.672 & -0.672 & 0.672 & -1.175 & 0.250 \\
\hline
-0.224 & -0.224 & -0.224 & -1.048 & 0.373 \\
\hline
-2.170 &  0.780 & 0.720 & -1.344 & 0.081  \\
\hline
\end{tabular}
\end{center}
\caption{ \label{X.2}
 The same as in Table 1 but $\sum \eta_i = -3\times 0.224$.}
\end{table}

\section{QSSR estimate of the slope of the IW function }
In view of the former result, let us estimate
$\xi'(1)$
{\it
analytically},
using QSSR in the Heavy Quark Effective Theory (HQET).
In so doing, we work as
\cite{DOS}-\cite{NEU},
with the ratio of vertex over two-point
function sum rules to two loops within the continuum model of \cite{BLO}.
Using the expressions in \cite{BAG}, \cite{NEU},
we can deduce the compact
analytical expression of the {\it physical} Isgur-Wise function:
\bea
\xi_{phys}(y \equiv vv') \simeq  \ga \frac{2}{1+y}\dr^2 \ga  1
+\frac{\alpha_s}{\pi} f(y) \dr \nnb \\
-<\bar qq>\tau^3 \ga \frac{8\pi^2}{3I_0}\dr \aga \ga 1+
\frac{\alpha_s}{\pi} 2.38 \dr \ga
1-\ga \frac{2}{1+y}\dr ^2 \dr +
\frac{\alpha_s}{\pi}g(y) \adr
\nnb \\
+<\alpha_s G^2> \tau^4 \ga \frac{8\pi^2}{3I_0} \dr
 \ga \frac{1}{192\pi}\dr (y^2-1) \nnb \\
+<\bar qq>\tau^5 \ga \frac{M_0^2}{4}\dr \ga \frac{8\pi^2}{3I_0}\dr
\aga \frac{2y+1}{3}-\ga \frac{2}{1+y}\dr^2 \adr,
\eea
with:
\bea
f(y)= \gamma(y)I_x(\tau) +(y-1)\ga \frac{16}{9}
\log 2
-\frac{49}{54} \dr -(1-y)^2 \ga \frac{8}{15}\log 2 -\frac{197}{600}\dr
\nnb \\
g(y)= \gamma(y)\ga Ei(-\omega_c \tau)-\gamma_E \dr
+(y-1)\ga \frac{16}{9}\log 2-\frac{56}{27} \dr
-(1-y)^2 \ga \frac{8}{15}\log 2-\frac{112}{225} \dr \nnb \\
\eea
where:
\bea
\gamma(y) =\frac{4}{3} \ga y \frac{
\log \ga y + \sqrt{y^2-1}\dr}{\sqrt{y^2-1}}
-1 \dr , \, \,
\gamma(1)=0 , \, \, \gamma'(1)=8/9, \nnb \\
 \gamma_E=0.5772, \, \,
Ei(-x)= -\int_x^{\infty} \frac{dx}{x}e^{-x}
\eea
and:
\beq
I_x(\tau) \equiv
{\int_{0}^{\omega_c} d\omega \frac{\omega^2}{1+\frac{\omega}{M_Q}}
 \log (\omega\tau)
e^{-\omega\tau}}/
\ga I_0 \equiv
\int_{0}^{\omega_c} d\omega \frac{\omega^2}{1+\frac{\omega}{M_Q}}
e^{-\omega\tau} \dr.
\eeq
We shall use in our numerical analysis the value: $\omega_c
\simeq (3.0 \pm 0.5)$ GeV with a generous error compared to the true
error of 0.1 \cite{SN1},\cite{ZAL}. We leave $\tau$ as a free parameter,
which we shall fix from a variational method.
The slope $\xi'(1)$ is the value of the first derivative of the IW
function with respect to
$y$ at $ y=1$, which we can deduce {\it analytically}
from the previous sum-rule expression of $\xi_{phys}$. Then, its
expression reads:
\beq
 \xi'(y=1)\equiv -1 +\delta_{pert}+\delta_{NP},
\eeq
where:
\bea
\delta_{pert} &\simeq & -\frac{\alpha_s}{\pi}(\tau)\ga
I_x \gamma'(1)+ \frac{16}{9}\log 2-\frac{49}{54}\dr \nnb \\
\delta_{NP} & \simeq &- \frac{8\pi^2}{3I_0}\tau^3
<\bar qq> \ga 1+1.05
\frac{\alpha_s}{\pi}-\frac{5}{12}M_0^2\tau^2 \dr \nnb \\
&+& \frac{8\pi^2}{3I_0}\tau^4 \frac {<\alpha_s G^2>}{96\pi}.
\eea
We optimize this previous sum rule for $\xi'$
using a variational method. The stability of the result is reached for
$\tau^{-1} \simeq 1.7$ GeV (which is about the characteristic scale of
the $b$ into $c$ transition), while the result is very stable
(less than $2\%$ change)
for a larger
range of $\tau^{-1}$ between 1.25 and 2.5 GeV.
An analogous stability is also obtained by moving $\omega_c$ in the range
between 2.7 GeV (starting of $\tau$ stability) to 3.5 GeV.
We also notice that there is an almost complete
cancellation between the perturbative radiative and non-perturbative
corrections where each strength does not exceed $4\%$. This feature
leads to
the accurate estimate:
\beq
\xi'(1) \simeq -(1.00 \pm 0.02).
\eeq
We can multiply the previous error by a factor two in order
to add a conservative
systematic error inherent in the method
and in the continuum model.

This value differs from the previous result in \cite{BAG} deduced
from a numerical polynomial two-parameter fit
of the Isgur-Wise function, where the errors given there
are quite doubtful.
Our result satisfies
the bound derived
previously and the one of \cite{TAR2}.
\section{Improved estimate of V$_{cb}$}

As an application of our result,
let us estimate the mixing angle $V_{cb}$  by using
the ARGUS \cite{ARG} and CLEO \cite{CLEO} data on $\xi(y)\times V_{cb}$.
We shall use in our numerical analysis the following parametrizations:
\bea
\xi(y) \simeq 1+\xi'(y-1) \nnb \\
\xi(y) \simeq \exp \aga \xi'(y-1) \adr \nnb \\
\xi(y) \simeq \aga (1+y)/2 \adr ^{2\xi'} \nnb \\
\xi(y) \simeq \ga \frac{2}{y+1} \dr
\exp \aga \ga 2\xi'+1 \dr \frac{y-1}{y+1}
\adr.
\eea
The two former are in line of the Taylor expansion used for y around 1.
The third is the pole parametrization and the fourth is based on overlap
integrals of meson wave functions
in a harmonic oscillator model.
 We shall normalize our result with
 the world average
$\tau_B \simeq
(1.48 \pm 0.10)$ ps of the {\it B}-lifetime given in \cite{DAN}:
\beq
V_{cb} \simeq  \ga \frac{1.48 \mbox{ps}}
{\tau_B} \dr^{1/2}\tilde V_{cb}.
\eeq
Given the previous value of the slope in (35),
we determine $V_{cb}$ from each data point and then, we
make a weighted average
of the different results.
{}From $B^0 \rar D^{*-} l \bar \nu$, we obtain in units of $10^3$:
\bea
\vert\tilde V_{cb}| & \simeq & 36.9 \pm 3.3  \, \, \, \,
\mbox{ARGUS 91} \nnb \\
&\simeq& 36.0 \pm 1.5 \, \, \, \, \mbox{CLEO 93},
\eea
while from $B^- \rar D^{*0} l \bar \nu$, we obtain in units of $10^3$:
\beq
|\tilde V_{cb}|   \simeq   40.6 \pm 2.3  \, \, \, \, \mbox{ARGUS 92}.
\eeq
Our results are the average of the ones from the
previous alternative parametrizations of $\xi$ in (36).
The last two parametrizations give almost
the same results of $V_{cb}$.
The errors given
there are
the largest ones from each parametrization and are only due to the data.
The choice of the parametrizations induces an extra error of 1.4.
Theoretical errors
induced by the ones of the slope are negligible. We take the average
of
previous results in (38) and (39).Then, we obtain the final best
estimate:
\beq
|\tilde V_{cb}|   \simeq   (37.3 \pm 1.2 \pm 1.4) \times 10^{-3},
\eeq
where we the first error
comes from the data. while the second one is
due to the different choices of parametrizations used in the literature.
We consider this result as
a noticeable improvement over the existing estimate of $V_{cb}$
(see e.g. \cite{DAN}, \cite{ALI}) thanks to a better control of the
value of the slope
from the QSSR estimate in (35).
This result also agrees with the lesser accurate
estimate (after rescaling
the lifetime used in \cite{SN4})
from the
$B$ into $D$,$D^*$ semi-leptonic decays within a QSSR estimate
of the form factors at zero momentum, where finite corrections due to
the $c$ and $b$ quark masses have been taken into account.

Moreover, a model-independent
result
from the phenomenological de Rafael-Taron-like bound in (27) gives:
\beq
|\tilde V_{cb}|   \leq    38.9 \times 10^{-3},
\eeq
which is enough strong for eliminating some
results given
in the literature.
\section{Conclusions}
We have estimated in Eq. (21) the sum of the $\Upsilon \bar BB$ couplings
using vertex sum rules. Using this information into the analysis of
\cite{TAR2},
we have derived in Eq. (27) a phenomenological bound on the slope of the
IW function. Finally,
we have reestimated in Eq. (35) this slope analytically from
QSSR. From the previous results, we have deduced the value of $V_{cb}$
in (40) and the bound in (41). The accuracy of this value of $V_{cb}$
is mainly do to the good control of the slope both from the sum rules
and from the de Rafael-Taron-like bound.

\section*{Acknowledgements}
I wish to thank
Josep Taron and Eduardo de Rafael
for numerous discussions and for reading the manuscript. I have also
enjoyed useful conversations with Ahmed Ali,
Hans Gunter Dosch and Thomas Mannel.
\noindent
\section*{Figure captions}
$\tau$-dependence of the $\Upsilon \bar BB $
coupling in the Laplace sum rule analysis.

\end{document}